WILEY-VCH

**Multiplexed Catheter-Integrated Pressure Sensing System for Endoluminal Interventions**

*Xiaotong Guo, Qindong Zheng, Jinshi Zhao, Bing Li\*, Eric Morgan Yeatman\**

X. Guo, E. M. Yeatman

Department of Electrical and Electronic Engineering, Imperial College London, London, SW7 2AZ, United Kingdom

Email: e.yeatman@imperial.ac.uk

Q. Zheng

Department of Bioengineering, Imperial College London, London, SW7 2AZ, United Kingdom

J. Zhao

Department of Metabolism, Digestion, and Reproduction, Imperial College London, London, SW7 2AZ, United Kingdom

B. Li

Institute for Materials Discovery, University College London, London, WC1E 7JE, United Kingdom

Email: bing.li@ucl.ac.uk

E. M. Yeatman

College of Science and Engineering, University of Glasgow, Glasgow, G12 8QQ, United Kingdom

Funding: UK Engineering and Physical Sciences Research Council (EPSRC, EP/X525649/1)

**Abstract**

Advances in flexible catheters pave the way for minimally invasive diagnosis and treatment of luminal organs and tubular structures through endoluminal interventions. A key challenge is in establishing non-constraining pressure monitoring at the interfaces between medical catheters and intraluminal anatomy exhibiting curvilinear contours, structural variability, and time-dependent physiological motion. This work presents a scalable and multi-purpose pressure sensing system for multidirectional monitoring of tissue interactions, establishing a robust solution for deploying diagnostic and therapeutic instruments in various types of endoluminal interventions. This approach provides an integrated system encompassing pressure sensors, catheters, and signal acquisition devices. A poly (vinylidene fluoride-co-trifluoroethylene) (P(VDF-TrFE)) film is miniaturized and configured into a multiplexed piezoelectric-based pressure sensor, providing flexibility and scalability in conforming to medical catheters with curved surfaces. The catheter is fabricated with a cost-effective and highly scalable fiber drawing technology, establishing a means of fast prototyping catheters with bespoke structures for sensor integration and medical instrument integration. The system achieves enhanced pressure detection sensitivity and a comparable sensing range, compared with state-of-the-art catheter-integrated sensors. Through in-vitro phantom studies, the system performs precise multi-directional sensing within various clinical endoluminal scenarios, showing its potential in digitalizing tissue interactions during endoluminal interventions.

## 1. Introduction

Flexible instruments have gained increasing popularity in many minimally invasive endoluminal interventions due to their advantages in navigating through tortuous passage [1]. The catheter-based intervention represent one of the most promising flexible instrument applications for diagnostic and therapeutic tools implementation covering endovascular (e.g., stent deployment), gastrointestinal (e.g., colonoscopy), and bronchial (e.g., bronchoscopy) interventions [2]. Manipulation of catheters during surgeries is normally assisted by image-guided technology. However, such manipulation shows an absence of intrabody monitoring of subtle pressure variations in the vicinity of catheter's contact with surrounding intralumenal structures. As a result, catheter interventions suffer from suboptimal interactions with surrounding tissues, reduced coupling efficiency, and unawareness of sudden increase in pressure [3]. Such drawbacks lead to a heightened risk of tissue damage and longer recovery time [4], and limited capability to detect small, unnatural protrusions within the complex natural of intraluminal anatomy, such as aneurysm [5].

Integrating catheters with sensitive, miniaturized and flexible pressure sensors can be a robust and novel approach to resolve these challenges [2a, 4, 6]. Emerging strategies in the miniaturization of soft material pressure sensors, encompassing capacitive [7], resistive [8], and piezoelectric [9] mechanisms, demonstrate considerable potential to advance progress in this context [10]. Recent advances in flexible sensor array have established versatile means in instrumenting balloon catheters for multiplexed tissue contact measurements on the balloon [2b, 11]. However, strategies for monitoring tissue interactions surrounding catheter bodies during navigation still remain limited. Many designs suffer from limited miniaturization and spatial resolution, leading to insufficient mapping for confined and complex intraluminal anatomy. Low sensitivity limits the measurement of subtle pressure, while high flexibility is needed for integration with millimeter-sized, tube-shaped catheters. Additionally, compatibility with commercial devices is another challenge, leading to difficulties in both electronic aspects (such as wiring connections) and mechanical aspects (such as sensor attachment).

Piezoelectric pressure sensor is a promising candidate to resolve the above challenges in sensitivity, flexibility, and miniaturization [12]. A high sensitivity is demonstrated in piezoelectric-based sensors due to their effectiveness in measuring subtle pressure changes while retaining a large measurement range [13], making them naturally suitable for intrabody pressure measurement. Flexibility in piezoelectric materials can be retained by the recent applications of copolymers or derivatives, such as poly(vinylidene fluoride)-co-trifluoroethylene (P(VDF-TrFE)) [14]. These novel piezoelectric materials exhibit lower elastic modulus than the conventional brittle substrates made of lead zirconate titanate [14b, 15]. Their integration with flexible substrates shows a simpler fabrication process and a higher conformity with soft geometry, making the piezoelectric-based sensors more suitable in integration with flexible instruments. Miniaturization of device made of piezoelectric materials can be achieved by sandwiching the active material between electrodes with methods such as spin-coating and evaporating deposition, which also presented as agile means in structing a web of sensing array [16].

In line with the pressure sensor design, the compatibility of the pressure sensor with catheters is essential to its practical use in clinical settings. However, designs of commercial catheters are usually compact and self-contained, making less room for the sensor integration [2a]. One solution is to fabricate a bespoke catheter with lumens for sensor coupling and accommodation for catheter-based instruments, as proposed in this paper. Thermal drawing is a novel, low-cost, and highly scalable technique for fabricating medical fiber with designated structure and versatile material. The process involves thermally softening an amorphous polymer-based, macro-structured fiber preform that has the same cross-sectional design as the target fiber [17]. The preform is then elongated into millimetre- or micrometre-scale fibers by applying an axial pulling pressure while maintaining the cross-sectional shape throughout the drawing process [18]. In recent decades, fibers produced using this technique have been studied for medical applications in sensing [19], actuation [20], robotics [21], and smart wearable devices [1b].

Herein, we present a multiplexed, piezoelectric-based, and catheter-integrated pressure sensing system for establishing a universal solution to enhance the tissue interaction monitoring during endoluminal interventions. A pressure sensor array is structured by sandwiching a thin film of P(VDF-TrFE) with two patterned electrode layers via deposition, yielding a scalable architecture that can be reconfigured and adapted in size. A bespoke surgical catheter is fabricated with the highly scalable and cost-effective thermal drawing technique, and it is constructed with multiple lumens for seamless sensor integration and medical instrument deployment. Leveraging the flexibility and miniaturization of the P(VDF-TrFE)-based sensor array, the sensor conforms intimately to the curved catheter surface and enables multi-directional pressure sensing during intraluminal navigation. A signal acquisition module within the system is tailored to intrinsic characteristics of the P(VDF-TrFE) sensor, facilitating real-time pressure recording with direct on-screen visualization. In general, our catheter-integrated pressure sensing system provides an integrated, scalable, and universal platform for monitoring the pressure interaction between tissue and surgical catheters, thereby laying the groundwork for the realization of smart endoluminal intervention systems in the near future.

## 2. Overview of Catheter-Integrated Pressure Sensing System

The catheter-integrated pressure sensing system is designed in a miniaturized scale to navigate through confined intraluminal anatomy and provide real-time pressure feedback visualization (**Figure 1**a,b). The integrated central working lumen allows deployment of various diagnostic and therapeutic devices, such as optical catheter, balloon catheter, and urinary catheter (Figure 1c). Through the adoption of a multilayer sandwich P(VDF-TrFE) thin-film fabrication technique, the sensor array is configured with a miniaturized size of 9 x 8 mm as illustrated in Figure 1d. The sandwich layout provides an ultra-thin cross-sectional profile (around 20 µm) and enables signal readout when external mechanical loadings applied (Figure 1e). The array is composed of three sensing elements (1.5 x 1.5 mm) wiring to their corresponding anode electrodes (1 x 1 mm), and a ground electrode (1 x 1 mm) is connected to all sensing elements. The electrodes are formed with one side of Au coating, while the sensing elements are formed by sandwiching the P(VDF-TrFE) thin film between two Au deposited layers. The wiring of all electrodes and sensing elements is built in the form of stretchable gold pattern (0.1 mm width) to provide a stretchability in axial and longitudinal directions.

Leveraging the flexibility and miniature design of P(VDF-TrFE) sensor, the fabricated sensor is conformed intimately to the curved geometry of medical catheters by wrapping over the distal end of the catheter body, as shown in Figure 1f. This sensor-catheter integration design enables multi-directional contact pressure monitoring for the navigation within confined intraluminal spaces. Given the optimized design of the bespoke catheter, conductive wires are embedded inside the catheter to avoid exposure to intraluminal environment. After soldering electrodes and external wires, the sensor array is then sandwiched by two protective layers (Polytetrafluoroethylene (PTFE) with a thickness of 0.1 mm).

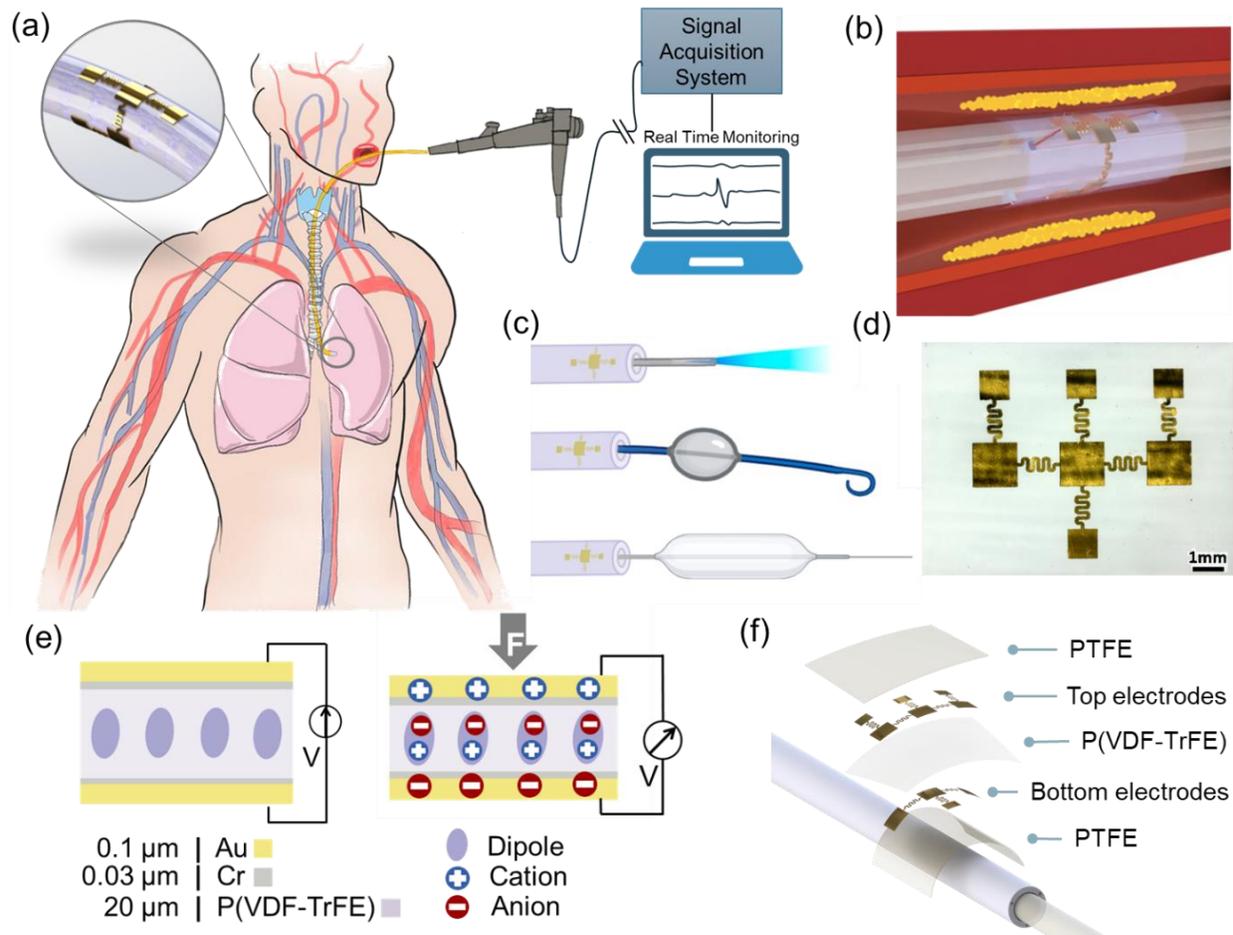

**Figure 1.** Illustration of the catheter-integrated sensing system. (a) Overview of the catheter-integrated pressure sensing system as applied in bronchial interventions. (b) Schematic of the catheter-integrated sensing system being in touch with an aneurysmal protrusion within a confined vascular structure. (c) Illustration of the catheter-integrated sensing system accommodating various medical devices, including optical catheter (top), urinary catheter (middle), and balloon catheter for stent deployment (bottom). (d) Microscopic view of the sensor array. (e) Cross-sectional schematics of the sensor design and piezoelectric mechanism showing the generation of potential difference when mechanical force is applied. (f) Demonstration of the device array in multilayer format integrated at the distal end of the catheter.

### 3. Design and Scalable Fabrication of the Multi-lumen Catheter

To enable seamless sensor integration and multifunctionality for versatile therapeutic and diagnostic instruments, the bespoke multi-lumen catheter was designed with a central hollow working lumen surrounded by multiple side lumens arranged around its circumference. The catheter has an outer diameter of 2.8 mm and a central lumen of 2.2 mm, allowing the passage of standard medical instruments up to 2.0 mm in diameter, such as endoscopes and guidewires. The cross-sectional design was optimized to support sensor configuration, with equilateral distribution of side lumens (each 100 μm in diameter) that facilitates organized routing of wires (80 μm in diameter) to connect anode and ground electrodes without overlapping or crossing.

For catheter fabrication, a scalable and low-cost fiber drawing process was used to produce polymeric fibers from three-dimensional (3D) printed preforms (**Figure 2**a). The preform material was selected as polycarbonate (PC) from various thermally drawable thermoplastic candidates (e.g., poly (methylmethacrylate) and acrylonitrile butadiene styrene) due to its good biocompatibility, high mechanical strength, and dimensional stability [21a]. The printed PC preform configures a 39.2 mm diameter, a 14-fold ratio to catheter's dimensions, and a length of 170 mm (Figure 2b). A continuous fiber exceeding 3 meters in length was fabricated from a single preform, maintaining a diameter tolerance within ±0.2 mm with the fiber drawing technique (Figure 2c). A 0.6-meter segment with consistent cross-sectional integrity was selected for use as the catheter body, providing sufficient length for endoluminal interventions.

Figure 2d shows the microscopic view of the cross-section of the resulting drawn fiber. The cross-section structure was well retained compared with the design of fiber geometry, allowing passing through the conductive wires therein (Figure 2e). The layout of the side lumens facilitates streamlined wiring of the integrated sensor that has electrodes uniformly distributed at 120° intervals. Figure 2f illustrates the stiffness characterization results of the fabricated fiber based on a flexural rigidity test. The fiber with embedded wires shows a stiffness of 1.8 N/mm for an identical 2.8 mm diameter and 50 mm length, according to the equation $Stiffness = Force/Displacement$.

**WILEY-VCH**

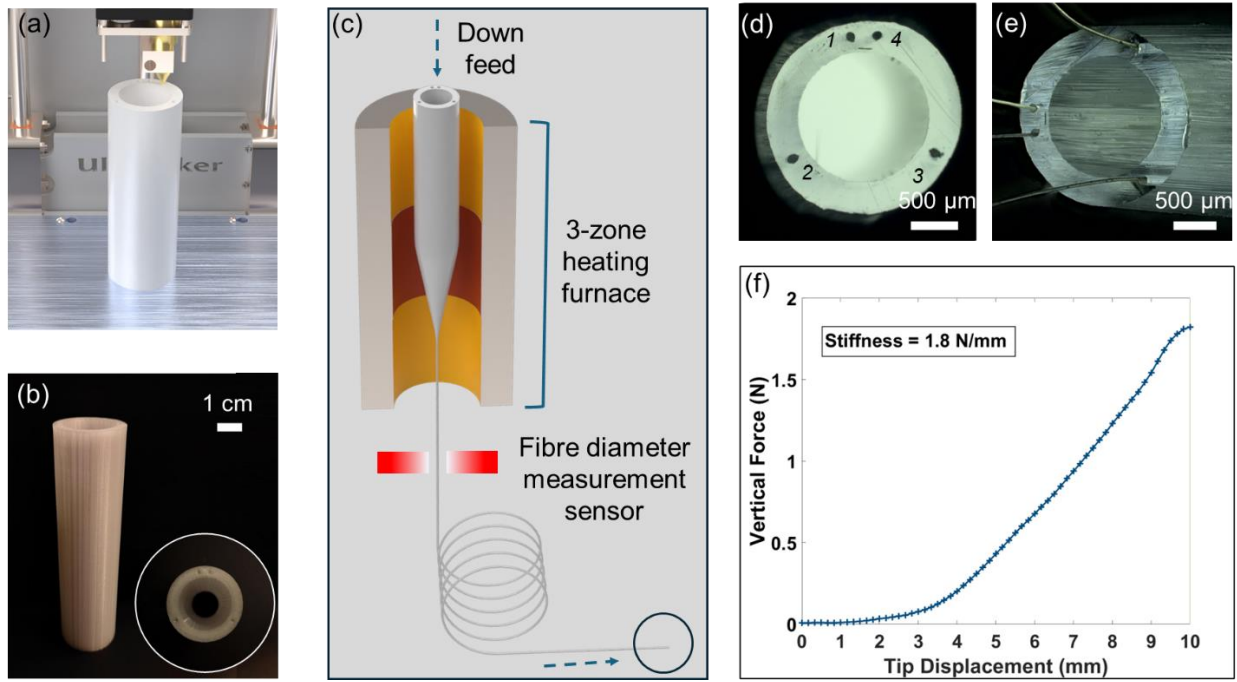

**Figure 2.** Catheter fabrication and characterization. (a) 3D printing of the preform with designated cross-section prepared for catheter fabrication. (b) Resulting 3D-print preform. (c) Schematic illustration of the fiber drawing process, where the colored hollowed cylinder represents the heating furnace. (d) Labeled microscopic cross-section view of the drawn fiber with channels 1-3 (the wiring channels for connecting anodes), 4 (the wiring channel for connecting cathode), and a central working channel for accommodating medical devices. (e) Microscopic view of the fiber with wires embedded. (f) Stiffness measurement of the fiber with a flexural rigidity test.

## 4. Signal Acquisition System

The impedance of piezoelectric-based sensor is significantly high with over 1 MΩ in low frequency [2a]. Recording of high-impedance sensors shows challenges in signal degradation, noise susceptibility, input impedance matching, amplification, and long-term drift. Thus, a dedicated signal acquisition system has been designed to record measured analogue signals from the sensor, transmit the signals to a computer following analogue-digital conversion, and achieve real-time monitoring of individual sensing units. As shown in **Figure 3**a,b, the system consists of four parts, sensor signal acquisition (SSA), analog signal processing (ASP), digital signal transmission (DST) and real-time display (RTD).

To test the signal acquisition system, different waveforms of variable frequencies generated from a function generator are fed into three input channels (C1, C2, and C3) of the circuit board in sequence, and each channel signal is plotted out over time correspondingly, as shown in Figure 3c. In detail, channel one first receives 1Hz, 5Hz and 10Hz sinusoidal signals in sequence while the other two channels remain near-zero baseline. Then square and ramp waveforms are received from channel two and three independently in turn. These results illustrate the circuit's ability to independently acquire signals from multiple channels without interference and visualize the results in a real-time display. Detailed structure of the system is illustrated in Figure 3d. Furthermore, the performance of our acquisition system exhibits close concordance with that of a commercial signal recording instrument (MSO-X 3054A) when measurements are conducted directly on a single sensing element of the piezoelectric sensor (Figure 3e).

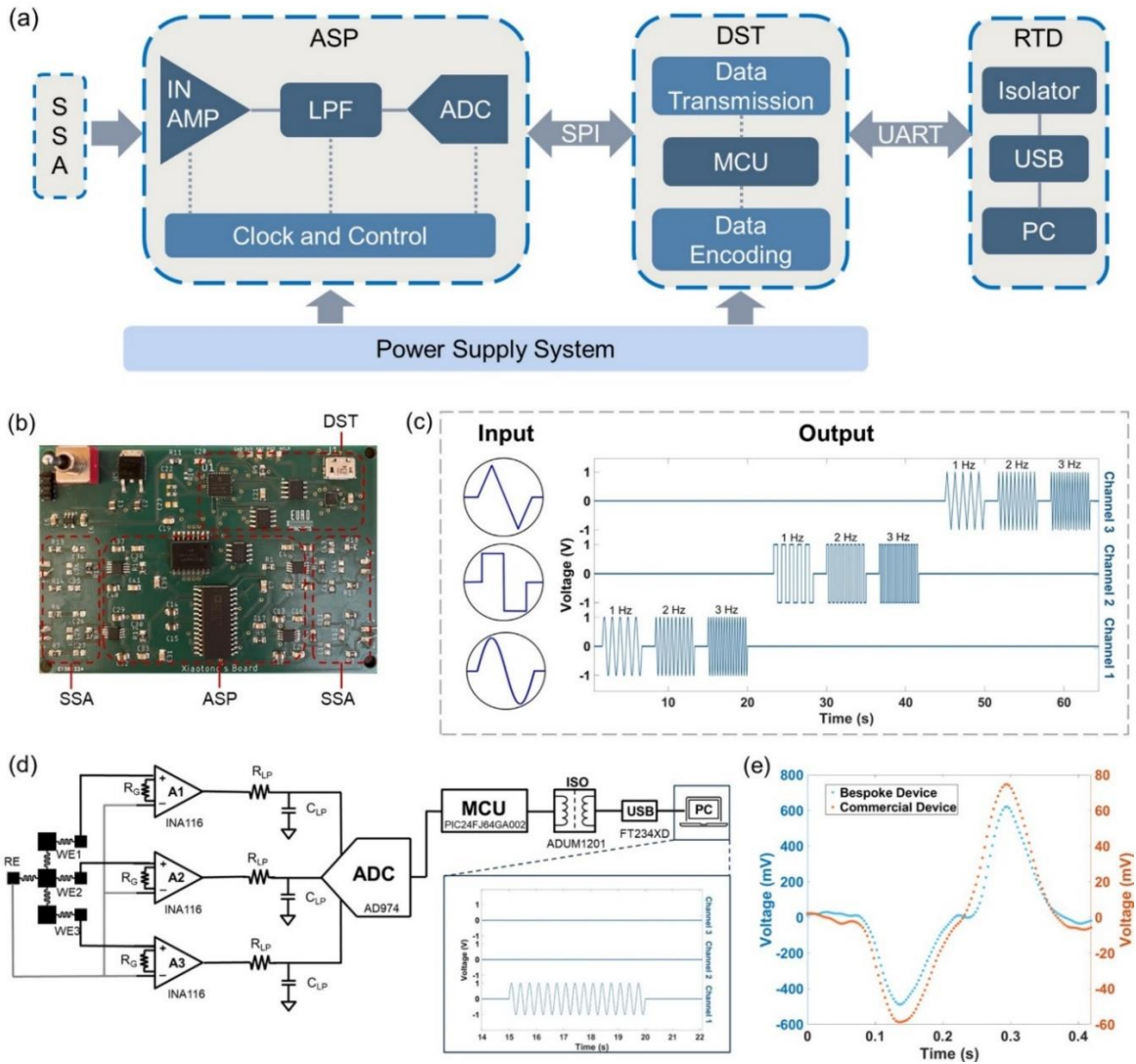

**Figure 3.** Design and characterizations of the signal acquisition system. (a) Schematic diagram of the signal acquisition and processing system for real-time sensor monitoring. (b) Image of the signal acquisition system board. (c) Signal readout characterization from three channels under different waveforms (ramp, square, and sinusoid) with different frequencies applied. (d) Circuit design: RE = reference electrode, WE = working electrode, $R_G$ = gain resistor of the instrumentation amplifier, $R_{LP}$ = low-pass filter resistor, $C_{LP}$ = low-pass filter capacitor, MCU=micro controller unit; and a graph showcases the processed digital output from the circuit. (e) Comparison of a commercial device with our signal acquisition system when measuring signals from our piezoelectric sensor.

## 5. Sensor Characterization

The impedance, sensitivity, voltage response, and multiplexing capability of this sensing system were characterized. The impedance of a single sensing unit was measured using an impedance analyzer (model E4990A, Keysight Technologies, CA) across a frequency range from DC to 100 kHz, as illustrated in **Figure** 4a. The sensing unit exhibited an impedance exceeding 10 MΩ at low frequencies, which is characteristic of a typical flat piezoelectric transducer and highlights its capacitive nature.

The sensing units and the multiplexed sensor array were characterized with our bespoke recording system. An experimental setup using a bespoke force application stage is demonstrated in Figure S1. As shown in Figure 4b, the sensor demonstrates a high linearity in voltage output across a typical range of intrabody pressures (0-80 kPa). This aligns with many catheter-based surgical procedures, for instance, catheter ablation, pressure from 0 to 26 kPa for a hemispherical catheter tip [22]. Its sensitivity ($S$) is determined by the equation $S=\delta V/\delta P$, where $V$ is the measured voltage at pressure $P$. Although the nature of high impedance, the sensor supported by our bespoke data acquisition system exhibits a 25 times higher sensitivity of 16 mV/kPa, compared with the state-of-the-art catheter integrated pressure sensors, as shown in Table S1. The voltage response is sensitive to the dynamic change of pressure (force loading and unloading) and demonstrates a negligible time delay under a loading rate of 0.5 kPa/ms, as shown in Figure 4c. The observed voltage response further demonstrates consistent and reproducible signal morphology across successive force-loading cycles under varying pressure conditions (Figure 4d), characterized by a distinct positive peak during loading and a corresponding negative peak during unloading. This behavior aligns with the principles of piezoelectric transducers subjected to dynamic mechanical stimuli, wherein the increase of pressure induces charge displacement toward the sensor's cathode, while the decrease of pressure generates a symmetric charge displacement in the opposite direction.

To evaluate the reproducibility in sensitivity, eight sensing arrays with a total of 24 sensing units were fabricated and individually characterized. The result presents an average of 16.1 mV/kPa and standard deviation of 0.7 mV/kPa under Gaussian approximation (Figure 4e). Uniformity of output signals in multiplexing was tested by sequential force application at a pressure of 10 kPa and loading cycle frequencies of 1 Hz on different sensing units, as shown in Figure 4f. Signals recorded in each 30 s long loading cycle reveal minor variance in morphology and amplitude. Negligible crosstalk was observed on the uncontacted channels. Noise was filtered with the built-in analogue and digital filters in our recording system.

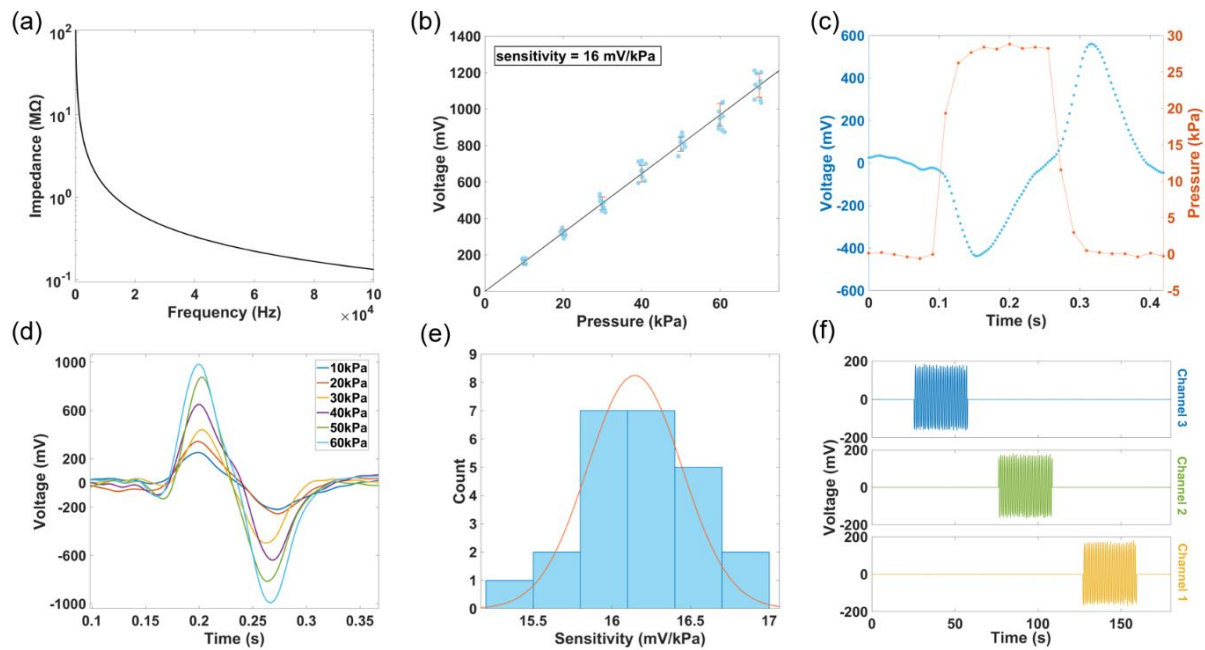

**Figure 4.** Characterization of the sensor array. (a) Semi-logarithmic plot demonstrating impedance variance of the sensor across different frequencies measured with commercial impedance analyzer. (b) Output peak voltage corresponding to various ranges of pressure. (c) Voltage response to a single cycle of force loading and unloading. (d) Voltage response corresponding to various ranges of pressure. (e) Histogram and approximated Gaussian line of the sensitivity measurements across

multiple sensing units. (f) Force application with a repeated loading and unloading cycle on the sensor array.

## 6. Medical Applications

The multiplexing performance of the sensor array upon assembling with the catheter was conducted on a 3D-printed model of a general biological conduit. To test the directional sensing of the device, the catheter was inserted into this conduit model where four pairs of protruded bumps were distributed with designated distances on the inner wall (**Figure** 5a). As the catheter navigated through the conduit, the sensor array measured pressure signals from two directions where the catheter body was in physical contact with the protruding bumps (Figure 5b). During the experiment, the delivery speed in catheter insertion at the proximal end remained constant at 15 mm/s with a motorized linear stage. Given the multi-directional pressure monitoring with the sensor array at the catheter's distal end, a reconstruction from the pressure recording can provide an estimation of contact locations. Figure 5c demonstrates the reconstruction of contact locations according to the recorded pressure peaks when the catheter passed each bump sequentially (Figure 5d), showcasing an accurate estimation with an average spatial uncertainty of 2.2 mm. Additionally, no crosstalk disturbance was observed at the uncontacted sensing unit.

Experiments in clinical scenarios with the integrated device were conducted using two types of in-vitro phantoms incorporating synthetic protruded tissues or tumors, representing an aortic arch, and a bronchial pathway. The selection of the phantoms was based on the typical catheter-based surgeries, i.e., endovascular and bronchial interventions. The in-vitro experiment on endovascular interventions started by navigating the catheter through the arch of aorta, with access via the descending aorta through an introducer. The catheter was then gently moved back and forth through the vessel multiple times Figure 5e,f. A distinct signal peak was observed at sensor 2, positioned on the catheter at the site of physical contact with the synthetic protrusion (Figure 5g). Smaller signal peaks were also recorded in the remaining sensors, likely attributable to contact between the catheter and the narrow vascular passage. This procedure and the sensor measurement were simultaneously recorded as shown in Supplementary Video S1. Figure 5(h) shows another set of measurements in which sensor 2 and sensor 3 were in contact with the protrusion. The complex shape of signals reveals the complicated and dynamic interactions between catheter and tissues.

An in-vitro study on a simplified bronchial phantom was conducted to illustrate the feasibility of the device in the clinical scenario of bronchial and lung interventions. Figure 5i,j showcases the catheter navigated through the right bronchi and passed a narrowed down air passage, representing bronchial inflammation or tumor. The sensor array recorded a significant negative spike of -0.6 V on the sensor in direct contact with the bump, representing an instant increase of pressure to around 40 kPa (Figure 5k). While the other sensing units in contact with the wall of bronchi were observed minor and discontinuous amplitude shifts. This showcases the sensor array can detect the direction of major dynamic physical contact while retaining the information of minor physical contact in other directions. The procedure and corresponding sensor measurements are shown in Supplementary Video S2.

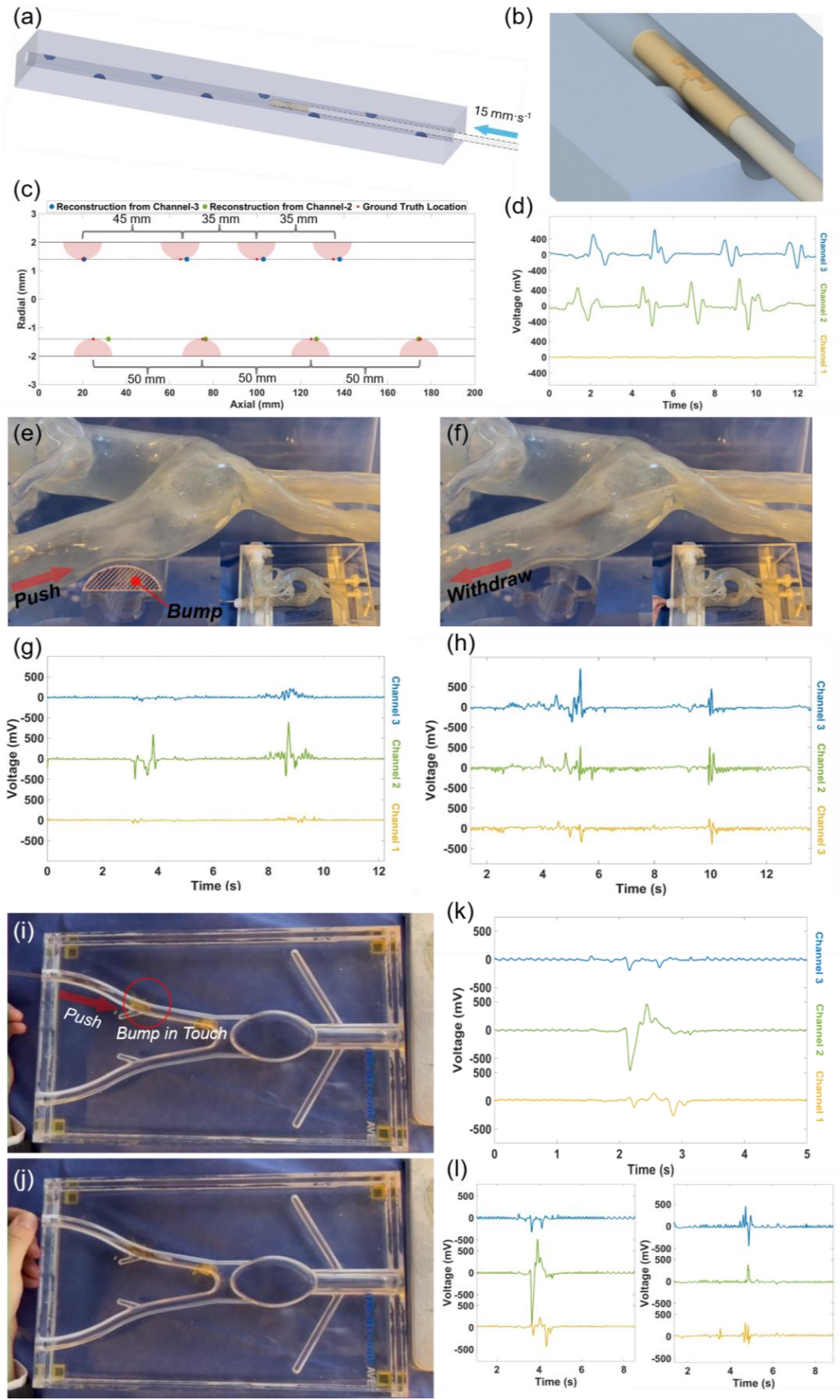

**Figure 5.** Results of in-vitro phantom studies. (a-b) Schematics of the biological conduit model displaying multiple organized protruding bumps. (c) Reconstruction of in-contact locations of each bump and comparison with ground truth labeling spacing distances. (d) Sensor output signals recorded during catheter insertion. (e, f) Image illustration of insertion and withdrawal of the catheter within an aortic arch phantom with an artificial bump. (g, h) A representative example of the sensor output signals recorded during the aortic arch phantom study. (i, j) Image illustration of catheter manipulation navigating through a simplified bronchial phantom with a minor protruding bump located at the midpoint. (k, l) Sensor array readout from representative examples.

## 7. Conclusion

In summary, a multiplexed catheter-integrated pressure sensing system has been developed for digitalizing tissue interaction during endoluminal interventions to minimize tissue damage and enhance surgical outcomes. The pressure sensing system consists of a miniaturized and flexible piezoelectric sensor array, a bespoke medical catheter hosting signal transmission and instrument deployment, and a signal acquisition module optimized for piezoelectric-based sensing materials. The sensor array is fabricated from a thin film of P(VDF-TrFE) with a highly configurable and miniaturized deposition process, while preserving the inherent mechanical flexibility of the material. To accommodate the electrical characteristics of the P(VDF-TrFE)-based sensor, the signal acquisition module is designed and optimized, achieving real-time signal readout and visualization. On top of it, the sensor demonstrates a 25 times higher sensitivity and a comparable sensing range from 0 kPa to 80 kPa, compared to the state-of-the-art catheter-integrated pressure sensors. The medical catheter platform is built to equip with multiple functional lumens and demonstrates seamless sensor integration and instrument deployment therein, facilitated by the highly scalable and low-cost thermal drawing technique. The integrated system exhibits high sensitivity in detecting pressure variations originating from multiple directions across the catheter surface. The in-vitro phantom studies further highlight the multi-directional sensing within confined anatomy. In general, the proposed multiplexed catheter-integrated pressure sensing system provides a scalable, fast-prototyping, and universal solution for the realization of smart endoluminal intervention systems in the near future.

## 8. Methods

### 8.1. Sensor-Catheter Assembly

The assembly began with cutting the sides of the fiber to expose the side lumens, gaining access for the conductive wires. The electrode pads for anodes and ground of the sensing elements were then connected with the pull-our wires via conductive Cu pads (**Figure S1**a). Subsequently, the wires were fed into their corresponding lumens until reaching the preassigned position (Figure S1b). After that, the sensor was gently bent and wrapped over the distal end of the fiber. The process was finalized by wrapping a layer of PTFE tape (100 μm thickness) over the entire sensor and corresponding wires, as a protective layer preventing sensors from scratch damage and short circuit caused by body fluids.

### 8.2. Fiber Drawing

The fiber preform was designed using 3D computer-aided design software (SolidWorks, Dassault Systèmes, France) and fabricated using a 3D printer (Ultimaker PC Transparent, 2.85 mm; Netherlands). This fiber preform was then placed in a vacuum oven at 70°C for 2 days to dehydrate and remove moisture content.

The fiber drawing process was initiated by securing the prepared preform onto a vertically aligned linear motorized stage positioned above a three-zone cylindrical heating furnace. Once the temperature at each zone of the furnace reaching 150°C, 230°C, and 85°C (from top to bottom), the linear stage continuously fed the fiber preform into the furnace at a constant speed of 1 mm/min ($V_{feed}$). After the polymer preform began to neck down, a downward pulling force was applied from the bottom side using a speed-controllable capstan ($V_{draw}$). The resulting diameter of the fiber ($D_{fiber}$) can be approximated using the following equation:

$$\frac{D_{fibre}^2}{V_{draw}} = \frac{D_{preform}^2}{V_{feed}} \quad (1)$$

Meanwhile, the diameter of the fiber was continuously monitored using a laser measurement device, which provides real-time feedback for the operator to adjust the capstan speed, thereby controlling the fiber diameter. Finally, fibers with desired diameter were cut into sections with the targeted length for catheter preparation.

### 8.3. Catheter Stiffness Characterization

The stiffness of the catheter was evaluated by conducting a flexural rigidity test. The distal end section, where the sensor mounted, was tested with wires embedded. The experiment involved vertically displacing the catheter tip (50 mm) by 10 mm deflection, repeated over three trials. A commercial force sensor was used to measure the overall force applied (Nano43, ATI Industrial Automation, USA). The stiffness is calculated with the total force required for different amplitudes of tip deflection and expressed in N/mm.

### 8.4. Sensor Array Characterization

The characterization of sensor array was performed on a bespoke platform for force loading and unloading with a controllable cycle frequency, as shown in **Figure S2**. The contact area was determined via a customized spring probe mechanism (**Figure S3**).

The sensitivity of individual sensors was tested by repeatedly applying cycles of various pressures. The effective pressure in each cycle was derived from equation $P=F/A$, where $F$ is the peak force from the scale and $A$ is the surface area of the probe tip. Pressures ranging from 10 kPa to 70 kPa with a step of 10 kPa were experimented. 10 repeated cycles were conducted for each pressure. The voltage output from the sensor was determined by the positive peak in each loading cycle. Due to the high linearity, a first-order polynomial fitting was used for sensitivity calculation.

### 8.5. Biological Conduit Test

The performance of the sensor array mounted on the catheter upon multiple directions was tested with a 3D-print biological conduit experiment setup (**Figure S4**). Catheter was delivered with only axis force for insertion, thereby maintaining a similar contact force when passing through each bump. The insertion speed of the catheter was maintained by fixing with a linear stage which is driven by a motor. Due to the distribution of the bumps, two sides of the sensing area should be in contact with the conduit while the remaining sensing area had no pressure applied.

### 8.6. In-vitro Phantom Study

Endovascular Interventions: a 1:1 soft silicone model of aortic arch with its entry of flow at 3-cusp (T-S-N-002-v2, Elastrat, Switzerland) was used in this phantom study, as shown in **Figure S5**. The experiment started from introducing the catheter into the phantom through a unidirectional valve, simulating the function of a catheter introducer in endovascular procedures. The catheter was then delivered back-and-forth to test its contact with the endovascular wall and the artificial bump.

Bronchial and Lung Interventions: a simplified model of bronchial air passages (Medtronic, US) was used, as shown in **Figure S6**. The experiment began with inserting the catheter in parallel with the bronchial conduit to navigate through the passages. The catheter was then moved along the conduit to allow physical contact with the protruded artificial bump.

**Acknowledgement**

The authors would like to thank Bruno Gil, Ruiqi Jiang and Zeyu Wang for help with experiment discussion. This work was funded by UK Engineering and Physical Sciences Research Council (EPSRC, EP/X525649/1).